\documentclass[preprint,amsmath,amssymb,aps,showkeys,showpacs,pra]{revtex4-1}
\usepackage[latin1]{inputenc}
\usepackage{graphicx}
\usepackage{dcolumn}
\usepackage{bm}
\usepackage{epsfig}
\usepackage{hyperref}
\usepackage{graphics}
\usepackage{subfigure}
\usepackage{rotating}
\usepackage{float}
\usepackage{color}
\usepackage{amsfonts}
\usepackage{amssymb}
\usepackage{latexsym}

\begin{document}

\title{ A Quantum Ring in a  Nanosphere  }

\author{A. L. Silva Netto}
\email{anibal.livramento@univasf.edu.br}
\affiliation{Colegiado de Engenharia Mec\^{a}nica, Universidade Federal do Vale do S\~{a}o Francisco, $48902-300$, Juazeiro, BA, Brazil}

\author{B. Farias, J. Carvalho}
\email{josevi@ccta.ufcg.edu.br}   
\affiliation{Unidade Acad\^{e}mica de Tecnologia de Alimentos, Centro de Ci\^{e}ncias e Tecnologia Agroalimentar, Universidade Federal de Campina Grande, 
Pereiros, $58840-000$, Pombal, PB, Brazil}

\author{C. Furtado}
\email{furtado@fisica.ufpb.br} 
\affiliation{Departamento de F\'{\i}sica, Universidade Federal da Para\'{\i}ba, Caixa Postal $5008$, $58051-970$, Jo\~{a}o Pessoa, PB, Brazil.}

\begin{abstract}
In this  paper we study the quantum dynamics of  an electron/hole in  a two-dimensional quantum ring  within a spherical space.  For this geometry,  we consider
a harmonic  confining potential.  Suggesting  that the quantum ring is  affected by  the presence of  an Aharonov-Bohm flux and an uniform magnetic field,   we
solve the Schr\"odinger equation for this problem and  obtain exactly the eigenvalues of energy  and corresponding eigenfunctions for this nanometric quantum system.  Afterwards, we calculate the
magnetization and persistent current are calculated, and  discuss  influence of curvature of space  on these values. 
\end{abstract}
\pacs{03.65.Ge, 68.65.Hb}
\keywords{Quantum Ring in curved space, Tan-Inkson confinement, Quantum ring in spherical geometry}

\maketitle
%
\section{Introduction}
In recent years, the study of confined quasiparticles in nanostructures  with annular geometry has attracted great interest in  condensed matter  physics. These quantum 
rings exhibit several interesting physical phenomena, such as the Aharonov-Bohm effect~\cite{9,10}, spin-orbit interaction effects~\cite{spin}, persistent 
currents~\cite{12,13}, quantum Hall effect~\cite{11} and the manifestation of Berry geometric quantum phase~\cite{14}.  More recent works have demonstrated that  assumption of finite 
width  brings intricacy from  an experimental point of view,  although even  more important results  have been also found. So, in Ref. \cite{lorke}  it was  shown  through use of
experiments with quantum rings of very small radii containing few electrons,  that there are some electron modes representing different radii of electronic orbits in 
these  nanometric systems. Several results show that  magnetic field penetration depth in  a conducting region plays an important role for physical properties of 
finite width quantum rings, including multiple channels  which were experimentally observed~\cite{12,13,inkson,margulis,bogachek}. One-dimensional quantum  rings pierced by Aharonov-Bohm
solenoid were used to observe the  quantum interference effect~\cite{spin,14,15,16}.  There are several exactly solvable models known for two-dimensional quantum 
rings, for example,  those ones  considered in refs.~\cite{12,inkson,margulis,bogachek}.  The theoretical approach  developed by Tan and Inkson~\cite{inkson,12} is of a special interest since it presents good agreement with experimental results, for instance, concerning effects of   magnetic field penetration in the conducting region.

Landau levels in negative~\cite{dune,comtet,comtet1} and positive~\cite{dune,greiter} curvature  cases have been intensively studied in order to explore quantum
Hall effect in these spaces~\cite{bulaphysb,takuya,jelal,iengo,nair,hasebe}. Quantum Hall effect in  a Lobachevsky plane was  considered in ref.~\cite{bulaphysb},
where the effect of negative curvature was observed  for a Hall conductivity of  these systems. The study of quantum Hall effect in  a spherical space was  carried out
for different scenarios in Refs.~\cite{hasebe2,nair2,nair3,nair3,nair4}. Advances in  the development of  techniques for low dimensional materials  motivate many 
investigations concerning curvature and topology influence on nanostructures, since  now it is possible to obtain several kinds  of curved two-dimensional 
surfaces~\cite{16} and  objects of nanometric size  with desired shapes~\cite{prinz}. The  impacts of  curvature and topology  for magnetic, spectral and 
transport properties of nanostructured materials have recently been studied by several authors~\cite{17,18,19,20}. The magnetic moment of  two-dimensional  electron gas  on  a negative  curvature surface was studied in ref. \cite{bulaemag}. The effect of  a negative curvature  for a quantum dot with impurity was investigated in ref.~\cite{geyler}. The zero mode in systems for spin-$1/2$ particle in the presence of an Aharonov-Bohm solenoid in Lobachevsky plane was  obtained in ref.~\cite{geyler1}. Recently, Bulaev, Geyler and Margulis ~\cite{bulacurva} have  studied  the Tan-Inkson  model  \cite{12,inkson} in  hyperbolic spaces and  provided  theoretical frameworks, comprising potentials with adjustable parameters, capable of describing nanostructures like quantum dots, antidots, rings and wires in this surface of negative curvature. Recently the effect of topology in quantum rings and dots was investigated in the refs.~\cite{22,23,lincoan}. 

In this work we study a nanosystem  in a positive curvature case. This  case  is interesting, among other reasons, because of the characteristics of the growth  techniques for nanometric systems,  such as quantum rings. Therefore, we  have  a motivation for probing how curvature influences physical properties of quantum rings. In our work we study a nanometric system grown over a surface with positive curvature, more specifically, a quantum ring in  a spherical space in the presence of  an Aharonov-Bohm flux and an uniform magnetic field through that space. We obtain the spectrum of energy and the wave function for Schr\"odinger equation solved exactly  for this system.  The magnetization  in the zero temperature case is obtained, and the influence of curvature  on  the magnetization is investigated. The persistent current 
is obtained using the Byers-Yang relation~\cite{byers}, and the influence of curvature on it  is discussed. An uniform magnetic field  in this case is introduced through 
the curved space in order to observe what happens in the conducting region. We also compare our results  in the appropriated limit with  results obtained in 
Ref.~\cite{bulacurva} for a ring on a  negative curvature surface. 

This paper  is organized as follows. In Section~\ref{sect2}  we investigate the quantum dynamics in  a  two-dimensional spherical space.  In Section~\ref{sec3},
we describe the  confinement potential for this positive curvature space. In Section~\ref{sec4} the  quantum dynamics of a charged particle confined in a 
Tan-Inkson  potential is investigated and  the eigenvalues and eigenvectors of energy  are obtained. In Section~\ref{sec5} the magnetization for $T=0$ is  found 
and the physical properties is discussed. The persistent current is  f calculated in Section~\ref{sec6}. Finally, in Section~\ref{sec7} we  present the concluding remarks. 

\section{Quantum dynamics  in a two-dimensional spherical space }\label{sect2}
\begin{figure}[!htb]
\includegraphics[width=\linewidth]{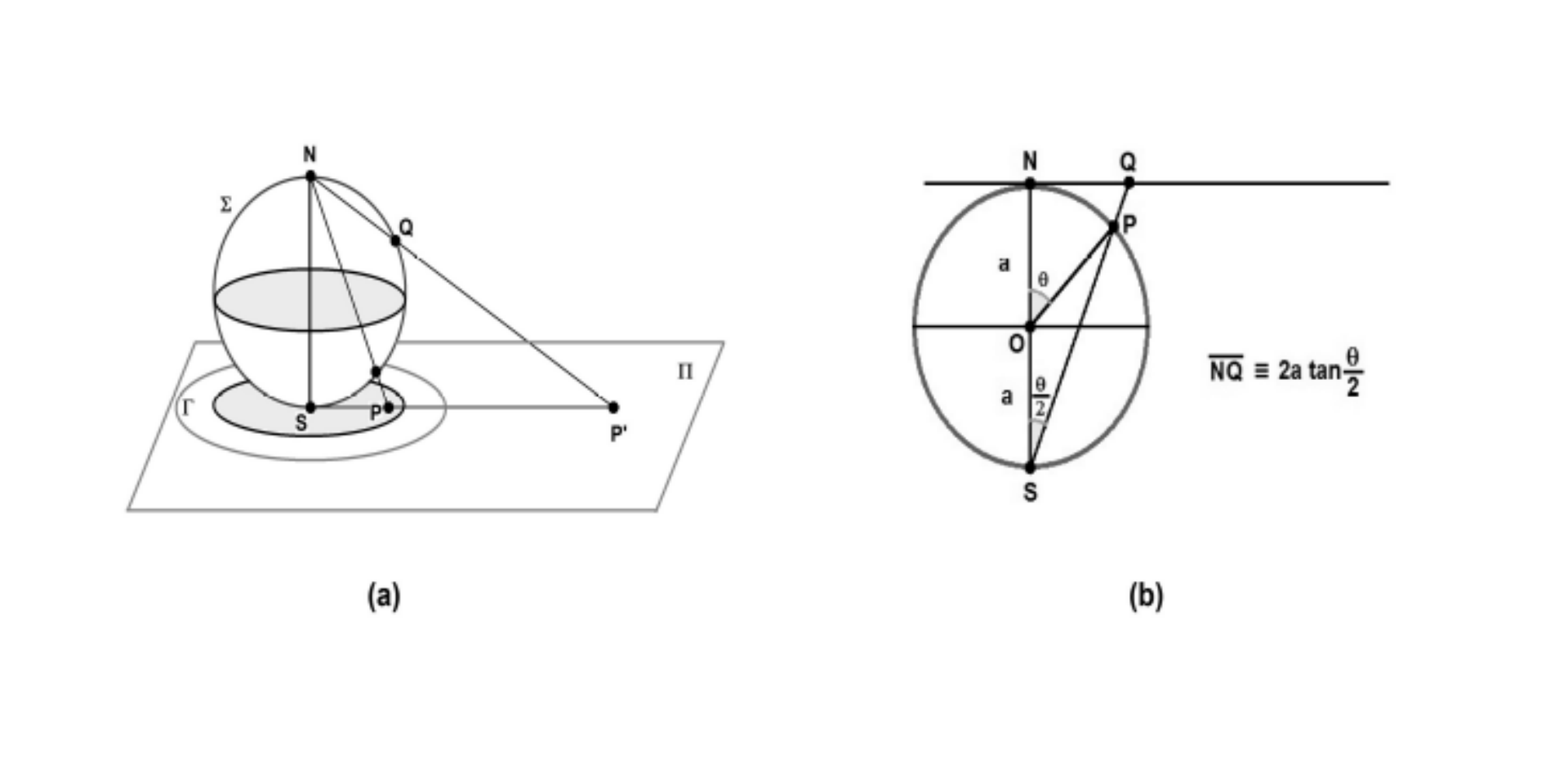}
\caption{(a) Stereographic projection of a sphere on a plane. (b) Trigonometric relation useful for obtaining the metric for the projected space.}
\label{fig:stereo}
\end{figure}

 First of all, we write the Hamiltonian for a free particle in a two-dimensional space $S^{2}$  described by a sphere embedded in  the Euclidean three-dimensional 
 space, $ x^{2}+ y^{2} + z^{2}=a^{2}$, where $a$ is the radius of sphere.  In this case the metric   on the sphere,  in terms of angular coordinates 
 $(\theta, \varphi)$ and sphere radius $a$, is given by
\begin{equation}
ds^{2}=a^{2}d\theta^{2}+a^{2} \sin^{2}\theta d\varphi^{2},
\label{sphMETRIC}
\end{equation}
where angular coordinates are restricted to the range $0 < \theta < \pi$ and $ 0 < \varphi < 2\pi$. In this study we use   a stereographic projection from the
points in a sphere with radius $a$   on a  plane. It is worth noting  that the  stereographic projection is a kind of map  preserving  angles and circles. 
In this way, first, angles between curves on original space   are mapped into equal  angles comprised by respective curves on projected plane; second, image of a 
circle on  the original space is also a circle on  the projected space. After this process, the points are  at the distance $\rho$ from the origin (that is, from the sphere's center)
on the projection plane. Here  the zenith angle is denoted by $\theta$ (which corresponds to the $\psi$ angle in the Figure ($\ref{fig:stereo}$). 
 In this way, we obtain the following relations: 
\begin{equation}
\tan\frac{\theta}{2} = \frac{\rho}{2a} \quad , 
\label{relation1}
\end{equation}
and
\begin{equation} d\theta^{2}=\frac{1}{a^{2}\left[1+\left(\frac{\rho}{2a}\right)^{2}\right]^{2}}d\rho^{2} \quad,
\label{relation2}
\end{equation}
and after some algebra we  find the metric describing our stereographically projected system:
\begin{equation}
ds^2=\frac{d\rho^2 +\rho^2 d\varphi^2}{\left[1+\left(\frac{\rho}{2a}\right)^2\right]^2} \ ,
\label{sphere-metric}
\end{equation}
where $0<\rho<\infty $ and $0< \varphi< 2\pi$.
We will consider an uniform  magnetic field $B$  on the spherical surface. So, for the projected representation, the equivalent magnetic field will be along the 
$z$-direction, perpendicular to the projection plane. The vector potential for this field configuration  for the geometry described by (\ref{sphere-metric}) is given by
\begin{equation}
\vec{A_{1}}=\left(0,\frac{B\rho}{2\left[1+\left(\frac{\rho}{2a}\right)^2\right]^2}\right)\label{uniformmag}.
\end{equation}
Now,  we introduce  a Aharonov-Bohm magnetic flux ($\Phi_{AB}$)~\cite{Landau:hydrogen3, sakurai, bogachek, Furtado:density, Dunne}  in $z$-direction  on the sphere. 
Therefore, the  only non-vanishing component of the magnetic vector potential is the azimuthal one,  and the corresponding vector potential is $\vec{A_{2}}$  given by 
\begin{equation}
\vec{A_{2}}=\left(0,\frac{\Phi_{AB}}{2\pi\rho}\right).\label{abflux}
\end{equation}
 The Hamiltonian in  the curved space  characterized by the metric  $g_{ij}$ in the presence of external magnetic fields is given by 
\begin{equation}
H_{0}=\frac{1}{2\mu \sqrt{g}} \left(-i\hbar \frac{\partial}{\partial
x^i}-\frac{e}{c}A_{i}\right)\sqrt{g}g^{ij} \left(-i\hbar \frac{\partial}{\partial
x^j}-\frac{e}{c}A_{j}\right). \label{curvedham}
\end{equation} 
Hence, the Hamiltonian (\ref{curvedham}) in  the  space with the metric (\ref{sphere-metric})  looks like
\begin{equation}
\begin{split}
\hat{H_{0}}\,=\,-\frac{\hbar^{2}}{2\mu a^{2}}\left\{a^{2}\left[1+\left(\frac{\rho}{2a}\right)^{2}\right]^{2}\left[\frac{1}{\rho}\frac{d}{d\rho}\left(\rho\frac{d}{d\rho}\right)+\frac{1}{\rho^{2}}\left(\frac{\partial}{\partial\varphi}+i\frac{\Phi_{AB}}{\Phi_{0}}\right)\right]\right\}\,
\\-\,i\frac{\hbar\omega_{c}}{2}\left[1+\left(\frac{\rho}{2a}\right)^{2}\right]\left(\frac{\partial}{\partial\varphi}+i\frac{\Phi_{AB}}{\Phi_{0}}\right)+\frac{\mu\omega_{c}^{2}\left(-a^{2}\right)}{2}\left(\frac{\rho}{2\left(ia\right)}\right)^{2}\,+\,\frac{\hbar^{2}}{8\mu a^{2}}\,.
\end{split}
\end{equation}
This operator  describes a quantum particle in a two-dimensional  spherical space submitted to an uniform magnetic field,  in the presence of Aharonov-Bohm solenoid in $z$-direction.

\section{The Tan-Inkson Con\-fi\-ne\-ment Po\-ten\-tial in a Two-\-di\-men\-si\-o\-nal Sphe\-ri\-cal Spa\-ce}\label{sec3}
  Let us introduce a confinement potential in a two-dimensional spherical space $S^{2}$. We generalize the  Tan-Inkson \cite{inkson} potential for this geometry.
This potential is a harmonic  confining potential  describing different  kinds of nanostructures in  a spherical space,  after  a simple change of parameters,  is given by
\begin{equation}
V(\rho)=\lambda_{1}\rho^2 + \frac{\lambda_{2}}{\rho^2}\left[1+\left(\frac{\rho}{2a}\right)^2\right]^2 - V_{0},\label{confpoten}
\end{equation}
where $\lambda_{1}$ and $\lambda_{2}$ are the parameters of the potential and $V_{0}$   looks like
\begin{equation}
V_{0}=\frac{\lambda_{2}}{2a^2}+2\sqrt{\lambda_{2}\left(\lambda_{1}+\frac{\lambda_{2}}{\left(2a\right)^4}\right)}.
\label{vo}
\end{equation}
The potential (\ref{confpoten}) has a minimum in $\rho_{0}$   equal to
\begin{equation}
\rho_{0}=\left(\frac{\lambda_{2}}{\lambda_{1}+\frac{\lambda_{2}}{(2a)^4}}\right)^{1/4}.
\label{minimum}
\end{equation}
It is worth noting  that, in the limit of $\lambda_{2} \to 0$  for the potential  presented by~(\ref{confpoten}), in stereographic coordinates, one recovers the 
harmonic potential for a quantum dot in  a flat space. Besides, if we consider  this system in the spherical  coordinates characterizing the metric (\ref{sphere-metric}), the quantum 
dot  takes the form $V(\theta)=4\lambda_{1}a^{2}\tan^{2}(\theta)$. For the  case $\lambda_{2} \to 0$, we  arrive at the antidot potential in spherical 
space.  In the limit $a \to \infty$, we obtain the flat Tan-Inkson potential in the form
\begin{equation}
V(\rho)=\lambda_{1}\rho^2 + \frac{\lambda_{2}}{\rho^2} - V_{0},\label{confpotenflat}
\end{equation}
where $V_{0}$ is given by
\begin{equation}
V_{0}=2\sqrt{\lambda_{2}\lambda_{1}}.
\label{vopla}
\end{equation}


\begin{figure}[!htb]
\includegraphics[width=0.40\textwidth]{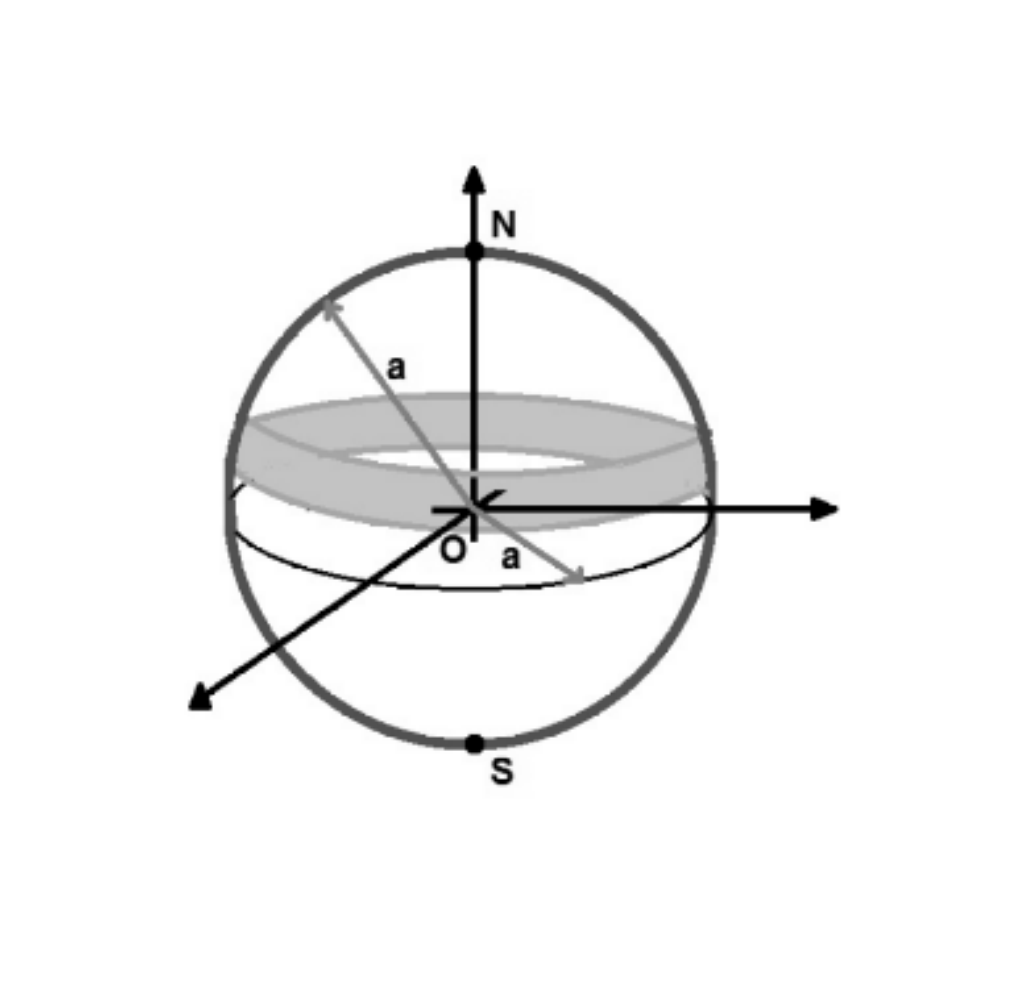}
\caption{Quantum ring on sphere}
\label{fig:dir}
\end{figure}
\section{The quantum Dynamics in a Quantum Ring in Spherical Space}\label{sec4}
 Now  we solve the Schr\"odinger equation for an electron/hole  confined  by the potential (\ref{confpoten}), in  the presence of magnetic fields 
 (\ref{uniformmag}) and (\ref{abflux}). In this  case the Hamiltonian of an electron  is  given by
\begin{equation}
\label{totalHamilt}
\begin{split}
\hat{H}\,=\,-\frac{\hbar^{2}}{2\mu a^{2}}\left\{a^{2}\left[1+\left(\frac{\rho}{2a}\right)^{2}\right]^{2}\left[\frac{1}{\rho}\frac{d}{d\rho}\left(\rho\frac{d}{d\rho}\right)+\frac{1}{\rho^{2}}\left(\frac{\partial}{\partial\varphi}+i\frac{\Phi_{AB}}{\Phi_{0}}\right)\right]\right\}\,
\\-\,i\frac{\hbar\omega_{c}}{2}\left[1+\left(\frac{\rho}{2a}\right)^{2}\right]\left(\frac{\partial}{\partial\varphi}+i\frac{\Phi_{AB}}{\Phi_{0}}\right)+\frac{\mu\omega_{c}^{2}\left(-a^{2}\right)}{2}\left(\frac{\rho}{2\left(ia\right)}\right)^{2}\,+\,\frac{\hbar^{2}}{8\mu a^{2}}\,
\\+\,\lambda_{1}\rho^{2}\,+\,\frac{\lambda_{2}}{\rho^{2}}\left[1+\left(\frac{\rho}{2a}\right)^{2}\right]^{2}\,-\,V_{0} \,.
\end{split}
\end{equation}
We  must solve the Schr\"odinger equation $\hat{H}\Psi(\rho,\varphi)=E\Psi(\rho,\varphi)$.  To do it, first we make  the following  change of the variable
\begin{equation}
x\equiv\frac{1}{1\,+\,\left(\frac{\rho}{2a}\right)^{2}}\,.
\label{def-x}
\end{equation}
Substituting this in (\ref{totalHamilt}), we obtain the following equation
\begin{equation}\label{schroequat}
\begin{split}
-\frac{\hbar^{2}}{2\mu a^{2}}\left\{a^{2}\frac{1}{x^{2}}\left[\frac{1}{a^{2}}\left(x^{2}\frac{d}{dx}\left[x(1-x)\right]\frac{d}{dx}\right)-\frac{1}{4 a^{2}}\left(\frac{x}{1-x}\right)\left(\frac{\partial}{\partial\varphi}+i\frac{\Phi_{AB}}{\Phi_{0}}\right)\right]\right\}\Psi(x,\varphi)\, 
\\ -\,i\frac{\hbar\omega_{c}}{2}\frac{1}{x}\left(\frac{\partial}{\partial\varphi}+i\frac{\Phi_{AB}}{\Phi_{0}}\right)\Psi(x,\varphi)+\frac{\mu\omega_{c}^{2}a^{2}}{2}\left(\frac{1-x}{x}\right)\Psi(x,\varphi)\,+\,\frac{\hbar^{2}}{8\mu a^{2}}\Psi(x,\varphi) \\ 
+\lambda_{1}\,4a^{2}\,\left(\frac{1-x}{x}\right)\Psi(x,\varphi) \,+\,\lambda_{2}\,\frac{1}{4a^{2}}\left(\frac{x}{1-x}\right)\frac{1}{x^{2}}\Psi(x,\varphi) \,-\frac{\lambda_{2}}{2a^{2}}\Psi(x,\varphi) - \\-\,\frac{\mu}{4}\omega_{0}^{2}\Psi(x,\varphi) \,\rho_{0}^{2}\, = E\Psi(x,\varphi).
\end{split}
\end{equation}
From~(\ref{schroequat}),  using the \textit{ansatz} $\Psi\left(\rho,\,\varphi\right)\,=\,\frac{e^{i\,m\varphi}}{\sqrt{2\pi}}\,f_{m}(x)$, we obtain the
following Schr\"odinger equation
\begin{equation}
\begin{split}
-\frac{\hbar^{2}}{2\mu a^{2}} \left\{\frac{1}{x^{2}}\left[x^{2}\,\frac{d}{dx}(x(1-x))\frac{d}{dx}\,-\,\frac{1}{4}\left(\frac{x}{1-x}\right)\left(m\,+\,\frac{\Phi_{AB}}{\Phi_{0}}\right)^{2}\right]\right\}\,f_{m}(x)\\
+\,\frac{\hbar\omega_{c}}{2}\left(m\,+\,\frac{\Phi_{AB}}{\Phi_{0}}\right)\,\frac{1}{x}\,f_{m}(x)\\
+\,\left\{\frac{\mu\,\omega_{c}^{2}\,a^{2}}{2}\frac{1}{x}\,-\,\frac{\mu\,\omega_{c}^{2}\,a^{2}}{2}\,+\,\frac{\hbar^{2}}{8\mu a^{2}}\,+\,\frac{\left[\frac{1}{2}\mu\,a^{2}\,\omega_{0}^{2}\,-\,\frac{1}{2}\,\mu\,a^{2}\,\omega_{0}^{2}\left(\frac{\rho_{0}}{2a}\right)^{4}\right]\,x}{1-x}\,+\,\frac{\frac{\mu\,a^{2}\,\omega_{0}^{2}}{2}}{x(1-x)}\right\}\,f_{m}(x)\\
-\left\{\,\frac{\left[\mu\,a^{2}\,\omega_{0}^{2}\,-\,\mu\,a^{2}\,\omega_{0}^{2}\left(\frac{\rho_{0}}{2a}\right)^{4}\right]}{1-x}\,+\,\mu\,a^{2}\,\omega_{0}^{2}\left(\frac{\rho_{0}}{2a}\right)^{4}\,+\,\frac{\mu}{4}\,\omega_{0}^{2}\,\rho_{0}^{2}\right\}\,f_{m}(x)=\,E\,f_{m}(x)\, ,
\end{split}
\label{sch1}
\end{equation}
where $\omega_{c}=\frac{eB}{m}$ is  a cyclotron frequency and 
\begin{equation}
\omega_{0} = \sqrt{\frac{8}{\mu}\left[\lambda_{1}+\frac{\lambda_{2}}{\left(2a\right)^4}\right]} \quad .
\end{equation}
Now, after some algebra we obtain 
\begin{equation}
\begin{split}
\left\{-\frac{d}{dx}(x(1-x))\frac{d}{dx}\,+\,\frac{M^{2}}{4}\frac{1}{1-x}\,+\,\frac{\mu^{2}a^{4}}{\hbar^{2}}\omega_{m}^{2}\frac{1}{x}\,-\,\frac{\mu^{2}a^{4}}{\hbar^{2}}\left[\omega_{c}^{2}\,+\,\omega_{0}^{2}\left(1\,+\,\left(\frac{\rho_{0}}{2a}\right)^{2}\right)^{2}\right]\right\}f_{m}(x)\\
=\,\left\{\frac{2\mu a^{2}}{\hbar^{2}}\,E\,-\,\frac{1}{4}\right\}f_{m}(x)
\label{schequatio1}
\end{split}
\end{equation}
where
\begin{equation}
M\equiv \sqrt{\left(m\,+\,\frac{\Phi_{AB}}{\Phi_{0}}\right)^{2}\,+\,\left(\frac{\mu\omega_{0}\rho_{0}^{2}}{2\hbar}\right)^{2}}\,,
\end{equation}

\begin{equation}
\omega_{m}\equiv\sqrt{\left(\omega_{c}\,+\,\frac{\hbar}{2\mu a^{2}}\left(m\,+\,\frac{\Phi_{AB}}{\Phi_{0}}\right)\right)^{2}\,+\,\omega_{0}^{2}},
\end{equation}

\begin{equation}
E'\,=\,\frac{2\mu a^{2}}{\hbar^{2}}\,E.
\end{equation}
Thus, rearranging terms in~(\ref{schequatio1}), we write
\begin{equation}
\begin{split}
\left\{\frac{d^{2}}{dx^{2}}\,+\,\left(\frac{1}{x}\,-\,\frac{1}{1-x}\right)\frac{d}{dx}\,+\,\left[\frac{-\frac{\mu^{2}a^{4}\omega_{m}^{2}}{\hbar^{2}}}{x}\,+\,\frac{-\frac{M^{2}}{4}}{1-x}\right]\frac{1}{x(1-x)}\right\}f_{m}(x)\\
-\,\left\{\frac{1}{4}-\frac{2\mu a^{2}}{\hbar^{2}}E\,-\,\frac{\mu^{2}a^{4}}{\hbar^{2}}\left[\omega_{c}^{2}\,+\,\omega_{0}^{2}\left(1\,+\,\left(\frac{\rho_{0}}{2a}\right)^{2}\right)^{2}\right]\right\}\frac{1}{x(1-x)}f_{m}(x)=\,0.
\label{sch5}
\end{split}
\end{equation}
The equation~(\ref{sch5})  replays the form of following differential equation
\begin{eqnarray}
&& \frac{d^{2}\,P(\xi)}{d\xi^{2}}\,+\,\left(\frac{1-\alpha-\alpha'}{\xi}\,-\,\frac{1-\gamma-\gamma'}{1-\xi}\right)\frac{d\,P(\xi)}{d\xi}\,+\,\left(\frac{\alpha\,\alpha'}{\xi}\,-\,\frac{\gamma\,\gamma'}{1-\xi}\,-\,\beta\,\beta'\right)\times\nonumber\\&\times& \frac{1}{\xi\left(\xi-1\right)}P(\xi)\,=\,0\,,
\label{hipergeo1}
\end{eqnarray}
where
\begin{equation}
\alpha\,+\alpha'\,+\,\beta\,+\,\beta'\,+\,\gamma\,+\gamma'\,=\,1\,.
\label{hipergeo-relat}
\end{equation}
It is important  to note that~(\ref{hipergeo1}) and~(\ref{hipergeo-relat}) are similar to  Eqs. ($5$) and ($5a$) in  Ref.~\cite{rubino}.
 Further, we can see that~(\ref{hipergeo1})    has the form of the  hypergeometric equation,  whose solution reads as
\begin{equation}
P(\xi)\,=\,\xi^{\alpha}\,\left(1-\xi\right)^{\gamma}\,F\left(a',b',c';\xi\right)\,,
\end{equation}
where
\begin{equation}
a'\,=\,\alpha\,+\,\beta\,+\,\gamma,\quad b'\,=\,\alpha\,+\,\beta'\,+\,\gamma,\quad c'\,=\,1\,+\,\alpha\,-\,\alpha'\,.
\end{equation}
Comparing~(\ref{hipergeo1}) and~(\ref{sch5}), we see that
\begin{equation}
1\,-\,\alpha\,-\,\alpha'\,=1\quad \to \alpha\,=\,-\alpha'.
\end{equation}
and at the  same way
\begin{equation}
\gamma\,=\,-\gamma'\quad\mbox{and}\quad \beta'\,=\,1\,-\,\beta\,.
\end{equation}
Let us define
\begin{equation}
\alpha\equiv \frac{\mu a^{2}}{\hbar}\,\omega_{m}\,,
\label{alfa-def}
\end{equation}
and
\begin{equation}
\gamma\equiv \frac{M}{2}\,.
\label{gamma-def}
\end{equation}
We can also use the relation
\begin{equation}
\beta\,\left(1\,-\,\beta\right)\,=\,\frac{1}{4}\,-\,\frac{2\mu a^{2}}{\hbar^{2}}E\,-\,\frac{\mu^{2} a^{4}}{\hbar^{2}}\left[\omega_{c}^{2}\,+\,\omega_{0}^{2}\left(1\,+\,\left(\frac{\rho_{0}}{2a}\right)^{2}\right)^{2}\right]\,.
\label{find-beta}
\end{equation}
From last relations it is  easy to see that we can assume
\begin{equation}
P(x)\,=\,x^{\alpha}\,\left(1\,-\,x\right)^{\gamma}\,F\left(\alpha\,+\,\beta\,+\,\gamma,\,\alpha\,+\,\gamma\,+\,1\,-\,\beta,\,1\,+\,2\alpha,\,x\right)\,.
\label{hipergeo2}
\end{equation}
From~(\ref{find-beta}) we find  the following expression
\begin{equation}
\beta\,=\,\frac{1}{2}\,\pm\,\sqrt{\frac{2\mu a^{2}}{\hbar^{2}}E\,+\,\frac{\mu^{2}a^{4}}{\hbar^{2}}\omega_{c}^{2}\,+\,\frac{\mu^{2}a^{4}}{\hbar^{2}}\omega_{0}^{2}\,+\,2\frac{\mu^{2}a^{4}}{\hbar^{2}}\omega_{0}^{2}\left(\frac{\rho_{0}}{2a}\right)^{2}\,+\,\frac{\mu^{2}a^{4}}{\hbar^{2}}\omega_{0}^{2}\left(\frac{\rho_{0}}{2a}\right)^{4}}.
\label{betas}
\end{equation}
Assuming, because of~(\ref{hipergeo2}), that
\begin{equation}
\alpha\,+\,\beta\,+\,\gamma\,\leqslant\,-n\,
\label{limit-n}
\end{equation} 
and considering~(\ref{alfa-def}), (\ref{gamma-def}) and (\ref{betas}), and solving condition~(\ref{limit-n}) when equality holds, 
 after some algebra  and  substituting the explicit values for $\alpha,\,\beta,\,\gamma,\,\omega_{m}^{2},\,M^{2}$,  basing on previously obtained relations, we obtain
 the following energy eigenvalues
\begin{equation}
\begin{split}
E\,=\,\frac{\hbar^{2}}{2\mu a^{2}}\left[\left(n\,+\,\frac{1}{2}\right)^{2}\,+\,\left(n\,+\,\frac{1}{2}\right)M\,+\,\frac{1}{2}\left(m\,+\,\frac{\Phi_{AB}}{\Phi_
{0}}\right)^{2}\right]\,+\,\hbar\,\omega_{m}\left(n\,+\,\frac{1}{2}\,+\,\frac{M}{2}\right)\\
+\,\hbar\,\omega_{c}\left(m\,+\,\frac{\Phi_{AB}}{\Phi_{0}}\right)\,-\,\frac{\mu\omega_{0}^{2}\rho_{0}^{2}}{4},
\end{split}
\label{stripe-energy}
\end{equation}
with $n\,\in\,\mathbb{N}:\,0\,\leqslant\,n\,<\,\frac{\mu\omega_{m} a^{2}}{\hbar}\,-\,\frac{M}{2}\,-\,\frac{1}{2}$\,. In the limit $\lambda_{1} =\lambda_{2}=0$ and $\Phi_{AB}=0$ we  find the results obtained by Dunne~\cite{Dunne} for Landau levels in  a  spherical space. Here we can see that eigenvalues are   only discrete,   in contrast with the  case of hyperbolic space  where the  Landau levels are  studied in Refs.~\cite{comtet, comtet1, Dunne} and the eigenvalues    can be discrete as well as and continuous.    In the limit  $a \to \infty$ we  recover the results obtained by Tan and Inkson \cite{12}   given by
\begin{eqnarray}\label{eingtanink}
E=\hbar\,\omega_{fm}\left(n\,+\,\frac{1}{2}\,+\,\frac{M}{2}\right) + \hbar\,\omega_{c}\left(m\,+\,\frac{\Phi_{AB}}{\Phi_{0}}\right)\, -V_{0}
\end{eqnarray}
where  flat definition $V_{0}$ is given by  Eq. (\ref{vopla}), and $\omega_{fm}= \sqrt{(\omega_{c}^{2}+\omega_{0})^{2}}$.
 
\section{The Magnetization for Quantum Ring in Spherical space}\label{sec5}
Considering  our system as a canonical ensemble,  one can obtain the magnetization~\cite{landau-stat}, from  a Helmholtz free energy and  a 
magnetic field,  as
\begin{equation}
\mathcal{M}(B)\,=\,-\left(\frac{\partial F}{\partial B}\right)_{T,N}\,=\,{\sum}_{n,m}\mathcal{M}_{n,m}\,f_{0}\left(E_{n,m}\right)\, 
\label{magA}
\end{equation}
where $N$ represents the number of electrons and $f_{0}$ is the Fermi distribution function.  In this case, magnetic moment for each $(n,m)$th state is given by
\begin{equation}
\mathcal{M}_{n,m}\,=\,-\frac{\partial E_{n,m}}{\partial B},
\label{mag3}
\end{equation}
with 
\begin{equation}
N\,=\,{\sum}_{n,m}\,f_{0}\left(E_{n,m}\right).
\label{magB}
\end{equation}

Noting that $\omega_{m}$ depends  on the magnetic field $B$, $\omega_{c}$,  one can write
\begin{equation}
\frac{\partial}{\partial B}\,=\,\frac{\partial\omega_{c}}{\partial B}\,\frac{\partial}{\partial\omega_{c}},
\label{derivaB}
\end{equation}
 which implies the following relation
\begin{equation}
\mathcal{M}_{n,m}\,=\,-\frac{e}{\mu c}\,\frac{\partial}{\partial\omega_{c}}\,E_{n,m}.
\label{mag-omegC}
\end{equation}
In this way, we can use the above relation and  the energy eigenvalues~($\ref{stripe-energy}$),   and  obtain
\begin{equation}
\mathcal{M}_{n,m}\,=\,\mu_{B}\frac{m_{0}}{\mu}\left[\hbar\left(2n+M+1\right)\frac{\partial}{\partial\omega_{c}}\omega_{m}\,+\,\left(m+\frac{\Phi_{AB}}{\Phi_{0}}\right)\right],
\label{comp-flat}
\end{equation}
where $m_{0}$ represents  the  electron rest mass and $\mu_{B}$ is Bohr magneton given by
\begin{equation}
\mu_{B}\,=\,\frac{e\,\hbar}{2 m_{0} c}.
\end{equation}
Taking into account that 
\begin{equation}
\frac{\partial}{\partial\omega_{c}}\omega_{m}\,=\,\frac{1}{\omega_{m}}\left[\omega_{c}\,+\,\frac{\hbar}{2\mu a^{2}}\left(m+\frac{\Phi_{AB}}{\Phi_{0}}\right)\right],
\end{equation}
finally we find
\begin{equation}
\frac{\mathcal{M}_{n,m}}{\mu_{B}}\,=\,-\frac{m_{0}}{\mu}\,\left[\left(2n\,+\,M\,+\,1\right)\frac{\omega_{c}\,+\,\frac{\hbar\left(m+\frac{\Phi_{AB}}{\Phi_{0}}\right)}{2\mu a^{2}}}{\omega_{m}}\,+\,m+\frac{\Phi_{AB}}{\Phi_{0}}\right].
\label{magneticmoment}
\end{equation} 
The expression (\ref{magneticmoment}) is the magnetization for $T=0$ for a two-dimensional  quantum ring in  a spherical space.
Applying limit $a \to \infty $, we recover the flat case~\cite{12,footnote1} for Gaussian CGS units. 

\section{The Persistent current in Quantum Ring in Spherical Space}\label{sec6}
In  this section we investigate the arising of  the persistent current in  a two-dimensional quantum ring for the spherical geometry.
Then, we also obtain persistent currents from ($\ref{stripe-energy}$).  It was showed  in \cite{byers-yang} that  for known energy eigenvalues, we can
obtain the current from the following relation
\begin{equation}
I_{n,m}\,=\,-c\,\frac{\partial E_{n,m}}{\partial\Phi_{AB}},\,
\end{equation}
that is, the Byers-Yang relation.
 In this way, the persistent currents will be given by
\begin{equation}
\begin{split}
I_{n,m}\,=\,-c\left\{\frac{\hbar^{2}}{2\mu a^{2}}\left[\left(n+\frac{1}{2}\right)\frac{\partial M}{\partial\Phi_{AB}}\,+\,\frac{1}{2}\frac{\partial}{\partial\Phi_{AB}}\,\left(m+\frac{\Phi_{AB}}{\Phi_{0}}\right)^{2}\right]\,+\,\hbar\left(n+\frac{1}{2}+\frac{M}{2}\right)\frac{\partial\omega_{m}}{\partial\Phi_{AB}}\right\}\\
\,-\frac{c}{2}\hbar\omega_{m}\frac{\partial M}{\partial\Phi_{AB}}\,-\,c\,\frac{\hbar\omega_{c}}{2}\frac{1}{\Phi_{0}}\,.
\label{persist-1}
\end{split}
\end{equation}

Taking into account that
\begin{equation}
\frac{\partial M}{\partial\Phi_{AB}}\,=\,\frac{1}{2}\,\frac{1}{M}\,2\,\left(m+\frac{\Phi_{AB}}{\Phi_{0}}\right)\,\frac{1}{\Phi_{0}}\,=\,\frac{1}{M}\left(m+\frac{\Phi_{AB}}{\Phi_{0}}\right)\,\frac{1}{\Phi_{0}}\,,
\label{diff-M}
\end{equation}
and
\begin{equation}
\frac{\partial\omega_{m}}{\partial\Phi_{AB}}\,=\,2\,\left(m+\frac{\Phi_{AB}}{\Phi_{0}}\right)\,\frac{1}{\Phi_{0}}\,=\,\frac{\partial\omega_{m}}{\partial\Phi_{AB}}\,=\,\frac{\hbar}{2\mu a^{2}}\,\frac{1}{\Phi_{0}}\,\left[\frac{\omega_{c}\,+\,\frac{\hbar}{2\mu a^{2}}\left(m+\frac{\Phi_{AB}}{\Phi_{0}}\right)}{\omega_{m}}\right]\,,
\label{diff-omegam}
\end{equation}
and substituting (\ref{diff-M}) and (\ref{diff-omegam}) in (\ref{persist-1}), one finds
\begin{equation}
\begin{split}
I_{n,m}\,=\,(-c)\left\{\left[\frac{\hbar^{2}}{2\mu a^{2}}\frac{\left(2n+1\right)}{2}\,+\,\frac{\hbar\omega_{m}}{2}\right]\frac{1}{\Phi_{0}}\frac{1}{M}\left(m+\frac{\Phi_{AB}}{\Phi_{0}}\right)\,+\,\frac{\hbar^{2}}{2\mu a^{2}}\left(m+\frac{\Phi_{AB}}{\Phi_{0}}\right)\frac{1}{\Phi_{0}} \right\}\\
-c\left\{\frac{\hbar}{2}\left(2n+M+1\right)\frac{\hbar}{2\mu a^{2}}\frac{1}{\Phi_{0}}\left[\frac{\omega_{c}\,+\,\frac{\hbar}{2\mu a^{2}}\left(m+\frac{\Phi_{AB}}{\Phi_{0}}\right)}{\omega_{m}}\right]\,+\,\frac{\hbar\omega_{c}}{2}\frac{1}{\Phi_{0}}\right\}\,.
\end{split}
\end{equation}
Here we used 
\begin{equation}
\Phi_{0}\,\equiv\,\frac{hc}{e}\,=\,\frac{2\pi\hbar c}{e}\,,
\end{equation}
where $e$ is the electron/hole  charge. After some algebraic manipulations we obtain
\begin{equation}\label{currentspherical}
I_{n,m}\,=\,\frac{c}{\pi\rho_{m}^{2}}\left\{\mathcal{M}_{n,m}\left[1+\left(\frac{\rho_{m}}{2a}\right)^{2}\right]\,+\,\mu_{B}\frac{m_{0}}{\mu}\frac{\omega_{c}}{\omega_{m}}\left(2n+1\right)\right\}\,,
\end{equation}
where
\begin{equation}
\rho_{m}\,\equiv\,\sqrt{\frac{2\hbar M}{\mu\omega_{m}}}\,,
\end{equation}
is the effective radius of the state with  a  quantum number $m$.


\section{Concluding remarks}\label{sec7}
In this  paper we have investigated the two-dimensional quantum ring in the presence of  the Aharonov-Bohm quantum flux and an uniform magnetic field in the 
spherical space. We  obtained the eigenvalues and eigenfunctions of  the Hamiltonian and  demonstrated the influence of curvature  on these physical quantities. We have  found the magnetization and  the persistent current for $T=0$ and  showed the influence of the curvature  on these cases. In the zero curvature limit ($a \to \infty$) we  reproduced the previous results obtained by Tan and Inkson \cite{12}. In the case  where $\lambda_{1}=\lambda_{2}=0$ and $\Phi_{AB}=0$ we  obtained the Landau levels in spherical space \cite{Dunne}.  Notice that  in the expression (\ref{currentspherical}) the  first contribution is  caused by the classical  current in a quantum ring of radius $\rho_{m}$  exposed to  a magnetic field, the second contribution  arises  due  to  the magnetic  field  penetration in  the   conduction region of the two-dimensional ring, and this contribution is responsible  for breaking the proportionality of the magnetic moment and the persistent current, a similar case was observed by Bulaev {\it et al.} for  a  two-dimensional quantum ring in hyperbolic space  \cite{bulaphysb}.  In the limit $\omega_{c} <<\omega_{0}$ the current is proportional to  the magnetic moment. We can write the magnetization in the following way:
\begin{equation}\label{magnetizationcurrent}
\mathcal{M}_{n,m}=\left\{ \frac{c\pi\rho_{m}^{2}}{\left[1+\left(\frac{\rho_{m}}{2a}\right)^{2}\right]} I_{n,m}\,-\,\mu_{B}\frac{m_{0}}{\mu}\frac{\omega_{c}}{\omega_{m}}\left(2n+1\right)\right\}\,.
\end{equation}
It follows from  Eq. (\ref{magnetizationcurrent})   that the magnetization can  be presented as  a sum  of  two terms. The first one arises  due to a magnetic dipole moment of a current loop  within spherical geometry,  and in the limit  $a \to \infty$ we  recover  the classical results ~\cite{jackson}. The another contribution  corresponds to a  diamagnetic shift.  This  term  has a contribution due  to the curvature  dependence in the term $\frac{\omega_{c}}{\omega_{m}}$. Finally, we claim that with the development on nanotechnology the possibility to investigate this spherical  system  from the experimental point of view can be realized  technically which can allow to  obtain this  spherical thin shell material . We  emphasize the interest  to   investigating the  spherical systems taken place in recent years,   see f.e  \cite{jelal,nair,hasebe,hasebe2,nair2,nair3}.

We can use the geometry of the quantum ring in spherical shell to construct an experimental set-up to investigate the physical properties obtained here for this theoretical model.  A nanometric system for a quantum ring on a sphere can be experimentally constructed between two well-determined $\theta$ angles in a quasi-two-dimensional nanostructured  hemisphere. Electron or holes may be injected by terminals attached to the ring and the persistent current can be measured in this experimental scheme similar to that described in reference Ref.\cite{gao} For flat case.  Other properties of these systems that can be investigated in a future publication, such as, de  Haas- van Alphen effect and a more detailed numerical study of  the contribution of the persistent current and magnetization obtained in previous sections to a system with many particles for $T = 0$  and $T \neq 0$.
\acknowledgments{This work was partially supported by  CNPq, CAPES  and FAPESQ.}

\end{document}